\documentclass[preprint,preprintnumbers,amsmath,amssymb]{revtex4}
\usepackage{amssymb,latexsym,amsmath}

\usepackage[activeacute,english]{babel}
\usepackage{fancyhdr}
\usepackage{textcomp}
\usepackage{amsfonts}
\usepackage{graphicx}
\usepackage{graphics}

\usepackage[active]{srcltx}

\newcommand{\al}{\alpha}

\newcommand{\om}{\omega}
\newcommand{\Om}{\Omega}

\newcommand{\rar}{\rightarrow}
\newcommand{\lrar}{\leftrightarrow}

\begin{document}

\title{Two charges on plane in a magnetic field: special trajectories}

\author{M.A.~Escobar-Ruiz}
\email{mauricio.escobar@nucleares.unam.mx}
\author{A.V.~Turbiner}
\email{turbiner@nucleares.unam.mx}
\affiliation{Instituto de Ciencias Nucleares, Universidad Nacional
Aut\'onoma de M\'exico, Apartado Postal 70-543, 04510 M\'exico,
D.F., Mexico}

\date{August 14, 2012}

\begin{abstract}
A classical mechanics of two Coulomb charges on a plane $(e_1, m_1)$ and $(e_2, m_2)$
subject to a constant magnetic field perpendicular to a plane is considered. Special "superintegrable" trajectories (circular and linear) for which the distance between charges remains unchanged are indicated as well as their respectful constants of motion. The number of the independent constants of motion for special trajectories is larger for generic ones. A classification of pairs of charges for which special trajectories occur is given. The special trajectories for three particular cases of two electrons,
(electron - positron), (electron - $\al$-particle) are described explicitly.

\end{abstract}


\maketitle

\begin{center}
\section*{Introduction}
\end{center}

It is well known that a classical charged particle placed to a constant uniform magnetic field performs a spiral motion: it moves with a constant velocity along magnetic field making a circular motion in a plane transversal to a magnetic field direction. Such a motion is characterized by a number of constants of motion, in particular, by pseudomomentum and by the third component of the canonical angular momentum directed along a magnetic line. Thus, the system on a transverse plane is superintegrable (see e.g. \cite{PW:2000}).
One can pose a question: would such a spiral trajectory continue to exist in the case when one more charged particle is present. The answer is affirmative although a number of those trajectories is dramatically reduced. They appear only if special initial conditions are chosen and not for any pair of charged particles. In this case the canonical angular momentum of the particle remains the constant of motion. Furthermore, in the presence of a second charge a new surprising trajectory may appear - the charged particle can move straightforwardly as a free particle. It seems natural that the second charge must perform a similar, circular (or straightforward) motion. They are characterized by the appearance of unusual, particular constants of motion - these are constants on a special trajectory only \cite{Turbiner:2012}. We call these trajectories special or superintegrable.

The goal of this paper is to describe pairs of charges as well as the initial data admitting the special trajectories. It is evident that the problem is reduced to a problem of two Coulomb charges placed on a plane subject to a magnetic field perpendicular to the plane. Some trajectories of this type for the case of two electrons and the electron-positron pair were found in \cite{CC:1997} and later in \cite{Taut:1999}.
Perhaps, it is worth noting that for a neutral system in $3D$ the motion is chaotic (see e.g. \cite{FW:1989}) and it remains true for a neutral system on the plane.
In general, for charges of opposite signs on the plane the system is not (completely)-integrable (see e.g. \cite{PM:2006}).

\bigskip

\section{Two charges in a magnetic field}

%\centerline{\bf Two charges in a magnetic field}

\bigskip

The Hamiltonian which describes two non-relativistic particles $(e_1,\, m_1)$ , $(e_2,\, m_2\,)$ placed on the plane subject to a constant and uniform magnetic field $\mathbf B=B\,\hat {\mathbf {z}}$ perpendicular to the plane has a form \footnote{$z$-axis is directed along magnetic field,  $c=1$\,. }
\begin{equation}
{\cal { H}} =   \frac{{({\mathbf {p}_1}-e_1\,{\mathbf A_1})}^2}{2\ m_1}
+ \frac{{({\mathbf {p}_2}-e_2\,{\mathbf A_2})}^2}{2\,m_2}
+ \frac{e_1\ e_2}{|{\boldsymbol \rho}_1 - {\boldsymbol \rho}_2 |}\ ,
\label{Hind}
\end{equation}
where ${\boldsymbol \rho}_{1,2}$ is a position vector of the first (second) particle, $\mathbf A_{1,2}=\frac{1}{2}\ (\mathbf B\times {\boldsymbol \rho}_{1,2})$. It is well known that the total Pseudomomentum
\begin{equation}
\begin{aligned}
&\boldsymbol K\ =\ \boldsymbol k_1 + \boldsymbol k_2\ =\ (\mathbf {p}_1+e_1\,\mathbf A_{1}) + (\mathbf {p}_2+e_2\,\mathbf A_{1})\ ,
\end{aligned}
\label{pseudomomentumIND}
\end{equation}
is a constant of motion, the Poisson bracket $\{\boldsymbol K,\,{\cal {H}} \}=0$, ${\boldsymbol k}_{1(2)}$ is pseudomomentum of the first (second) particle. The total \emph{canonical} momentum ${\boldsymbol L}^{total}_z$ is perpendicular to the plane
\begin{equation}
\begin{aligned}
\boldsymbol L^{total}_z&\ =\ {\boldsymbol \rho}_1\times {\bf p}_1+ {\boldsymbol \rho}_2\times {\mathbf p}_2\,,
\end{aligned}
\label{LzT}
\end{equation}
and is also conserved, $\{ \boldsymbol L^{total}_z, {\cal {H}} \}=0$. Hence, the problem is characterized by three conserved quantities (integrals) $K_{x,y},\ L_z^{total}$, where some $(x,y)$-coordinate system is introduced on the plane. Hence, in general, the problem (\ref{Hind}) is integrable, a number of integrals (including the Hamiltonian) is equal to the dimension of configuration space. Since the integrals $K_{x,y},\ L_z^{total}$
are not in involution (see below) the problem is not completely-integrable. These integrals are of a global nature: they are constants on any trajectory.

\bigskip

Let us introduce c.m.s (canonical) variables in a standard way
\begin{equation}
\begin{aligned}
&\mathbf R = \mu_1\, {\boldsymbol \rho}_1 + \mu_2\,{\boldsymbol \rho}_2 \ ,
\quad  {\boldsymbol \rho}= {\boldsymbol \rho}_1 - {\boldsymbol \rho}_2\ ,
\\ & \mathbf {P} = {\mathbf p}_1 + {\mathbf p}_2 \ ,
\qquad {\mathbf p} = \mu_2\,{\mathbf p}_1 -  \mu_1\,{\mathbf p}_2\ ,
\end{aligned}
\label{CMvar}
\end{equation}
where $\mu_i=\frac{m_i}{M}$  is a reduced mass and $M = m_1 + m_2$ is the total mass of the system. In these coordinates the total pseudomomentum
\begin{equation}
\boldsymbol K  = \mathbf P+q\,\mathbf A_{\mathbf R} + e_c\,\mathbf A_{{\boldsymbol \rho}}  \ ,
\label{pseudomomentum}
\end{equation}
where $q = e_1 + e_2$ is the total (net) charge and $$e_c = (e_1\,\mu_2-e_2\,\mu_1)\,,$$ is the so-called {\it coupling} charge,
\[
   {\bf A}_{\bf R}\ = \ \frac{1}{2}({\bf B} \times {\bf R})\ ,\
{\bf A}_{\boldsymbol \rho}\ = \ \frac{1}{2}({\bf B} \times {\boldsymbol \rho})\ ,
\]
and
\begin{equation}
  \boldsymbol L^{total}_z  = ({\mathbf R}\times {\mathbf P}) + ({\boldsymbol \rho}\times \mathbf p)\equiv \boldsymbol L_z + \boldsymbol \ell_z \ ,
\label{LzT2}
\end{equation}
where $\boldsymbol L_z, \boldsymbol \ell_z$ are cms and relative canonical angular momenta, respectively.

The integrals
$\boldsymbol K, \boldsymbol L^{total}_z$ obey the commutation relations
\begin{equation}
\begin{aligned}
&\{ K_x,\,K_y \} = -q\,B\ ,
\\ & \{ L^{total}_z ,\,K_x \} = K_y\ ,
\\ & \{ L^{total}_z ,\,K_y \} = -K_x\ ,
\label{AlgebraInt}
\end{aligned}
\end{equation}
with the Casimir operator ${\cal C}$ given by
\begin{equation}
 {\cal C}\ =\ K_x^2+K_y^2-2\,q\,B\ L^{total}_z \ .
\label{Casimir}
\end{equation}
Evidently, ${\cal C}$ is the global integral. For the case of a single charged particle ${\cal C}$  is, in fact, the Hamiltonian.

\hskip 1cm
It is convenient to make a canonical transformation which leaves the coordinates unchanged \cite{GD:1967},
\begin{equation}
\begin{aligned}
& {\mathbf {P}}^{\prime} = {\mathbf P} + e_c \, \mathbf A_{{\boldsymbol \rho}}\,,
\qquad{\mathbf {p}}^{\prime} = {\mathbf p} - e_c\,\mathbf A_{\mathbf R}\,,
\\ & \mathbf R^{\prime} = \mathbf R\,, \qquad \qquad \quad  \, \, \boldsymbol \rho^{\prime} = \boldsymbol \rho\ .
\label{CanTrans}
\end{aligned}
\end{equation}
In these coordinates the Hamiltonian (\ref{Hind}) takes the form
\begin{equation}
\begin{aligned}
{\cal {H}} &=  \bigg[\frac{ {( \mathbf {P}^{\prime}-q\,\mathbf A_{\mathbf R^{\prime}}-2\,e_c\,\mathbf A_{{\boldsymbol \rho^\prime}} )}^2}{2\,M}\bigg]
+ \bigg[\frac{{({\mathbf {p}}^{\prime}-q_\text{w}\,{\mathbf A_{{\boldsymbol \rho^\prime}}})}^2}{2\,m_{r}}
+ \frac{e_1\,e_2}{\rho^\prime}\bigg]\\
   & \equiv {\cal {H}}_{R}(\mathbf {P},\mathbf R)+{\cal {H}}_{\rho}
   (\mathbf {p},\boldsymbol \rho) ,
\label{HC}
\end{aligned}
\end{equation}
where $$q_{\rm{w}} \equiv e_1\,\mu_2^2+e_2\,\mu_1^2\,,$$ is an effective charge (the
weighted total charge) and $m_r=\frac{m_1\,m_2}{M}$ is the reduced mass. The case $e_c=0$
corresponds to a separation of cms motion from the relative one, the transformation (\ref{CanTrans}) becomes identical, a relation
\begin{equation}
\label{K2}
\frac{\boldsymbol {K }^2}{2\,M}\ =\ B \,\frac{e_1}{m_1}L_z\ +\ {\cal {H}}_{R}  \ ,
\end{equation}
occurs. The total pseudomomentum coincides to the cms pseudomomentum, see (\ref{pseudomomentum}). In the case $q=0$ the cms momentum is conserved, $\{\mathbf {P}^{\prime} , H\} = 0$. It reflects the fact that the center-of-mass motion is a motion of a free (composite) particle.

The Newton equations read
\begin{equation}
\begin{aligned}
& m_1\,{\ddot{\boldsymbol \rho}}_1 = e_1\,\dot{\boldsymbol \rho}_1\times \mathbf B + \frac{e_1\,e_2}{\rho^3}\ ({\boldsymbol \rho}_1-{\boldsymbol \rho}_2)\ ,
\qquad m_2\,\ddot{\boldsymbol \rho}_2 = e_2\,\dot{\boldsymbol \rho}_2\times \mathbf B - \frac{e_1\,e_2}{\rho^3}\ ({\boldsymbol \rho}_1-{\boldsymbol \rho}_2)\ .
\end{aligned}
\label{NEsep}
\end{equation}
In c.m.s variables (\ref{CMvar}) the equations of motion are reduced to
\begin{equation}
\begin{aligned}
&  M\,\ddot {\mathbf R} = q\,\dot{\mathbf R}\times \mathbf B + e_c\,\dot{\boldsymbol \rho}\times \mathbf B\,,
\\ & m_{r} \,\ddot {\boldsymbol \rho} = q_{\rm{w}}\,\dot{\boldsymbol \rho}\times \mathbf B - \frac{e_c^2\,B^2}{M}\,\boldsymbol \rho-\frac{q\, e_c\, B^2}{M}{\mathbf {R}} +\frac{e_c}{M}(\boldsymbol K\times \mathbf B)+ \frac{e_1\,e_2}{\rho^3}{\boldsymbol {\rho}}\,.
\end{aligned}
\label{Eq}
\end{equation}

It has to be emphasized the existence of two important particular cases,

\bigskip

(i)\ $q=0$ (neutral system) where the components of $\boldsymbol {K}$ are in involution, $\{ K_x, K_y \}=0$\ ,

\bigskip

(ii)\ $e_c=0$ for which in c.m.s. coordinates in the equations (\ref{Eq}) the variables are separated, two extra integrals appear:
${\cal {H}}_{\rho}(\mathbf {p},\boldsymbol \rho)$ and $\boldsymbol \ell_z$ (cf.
(\ref{pseudomomentum}), (\ref{LzT2}))\ . The total number of functionally-independent integrals is equal to four, $(K_{x,y}, {\cal {H}}_{\rho}, \ell_z)$. The integrals (${\cal {H}}_{R},\,{\cal {H}}_{\rho},\,L_z,\,\ell_z$) are in involution.

\bigskip

By balancing the Coulomb and Lorentz forces acting on the first (second) particle with a possible centripetal force it can be shown that
if the initial conditions for (\ref{NEsep}) are chosen in such a way that the initial velocities ${{\bf v}_{1,2}}$ are collinear and ${\rm v}_1 > {\rm v}_2$, and the vector of the relative distance between particles ${\boldsymbol \rho}$ is perpendicular to the velocities,
\begin{equation}
{\boldsymbol \rho}_{1}-{\boldsymbol \rho}_{2} = \frac{(\rm{v}_1-\rm{v}_2)}{\om}\,\mathbf{\hat x}\ ,
\qquad \dot {\boldsymbol \rho}_{1} = -{\rm{v}_1}\,\mathbf{\hat y}\ ,
\qquad \dot {\boldsymbol \rho}_{2} = -{\rm{v}_2}\,\mathbf{\hat y}\ ,
\label{constrain1}
\end{equation}
constrained to
\begin{equation}
  {M\,({\rm v}_1-{\rm v}_2)}^2\,(m_1\,{\rm v}_1+m_2\,{\rm v}_2)\,{\rm v}_1\,{\rm v}_2\,e_c-e_1\,e_2\,B\,{(e_1\,{\rm v}_1+e_2\,{\rm v}_2)}^2=0\ ,
\label{constrain2}
\end{equation}
where
\[
  \om = \frac{B\,(e_1\,{\rm v}_1+e_2\,{\rm v}_2)}{m_1\,{\rm v}_1+m_2\,{\rm v}_2} > 0\ ,
\qquad B\,{\rm v}_1 \, {\rm v}_2 \, e_1\, e_2\, e_c \neq 0 \ ,
\]
the (special) circular trajectory occurs with \footnote{It is implied that ${\boldsymbol \rho}_{1,2}$ are measured from the center of the circle\,.}
\begin{equation}
{\boldsymbol \rho}_{1,2}\ =\ \frac{{\rm v}_{1,2}}{\om}\,(\cos \om t,\, -\sin \om t)\ .
\label{traject}
\end{equation}
For that trajectory the distance between charges $|{\boldsymbol \rho}_{1}-{\boldsymbol \rho}_{2}|$ remains constant.

It is worth noting that the equations (\ref{NEsep}) (or (\ref{Eq}))
with the initial conditions (\ref{constrain1})-(\ref{constrain2}) are invariant:

$\bullet$  For fixed magnetic field $B$

\[
 (e_{1,2}\rightarrow - e_{1,2}\ ,\ \rm v_{1,2}\rightarrow  -\rm v_{1,2})
\]

$\bullet$  For fixed charges

\[
( B \rightarrow  -B\ ,\ {\rm v_{1,2}} \rightarrow  -{\rm v_{1,2}} )
\]

$\bullet$  For fixed initial velocities

\[
( B \rightarrow  -B\ ,\ e_{1,2} \rightarrow  -e_{1,2})
\]

The corresponding trajectories remain unchanged under these transformations.

%\newpage

\section{ Special Trajectories}

%\begin{center}
\subsection{Configuration I\ ($\rm{v}_1=-\rm{v}_2$)}
%\end{center}

\bigskip

\begin{figure}
\begin{center}
\includegraphics[width=2.5in,angle=0]{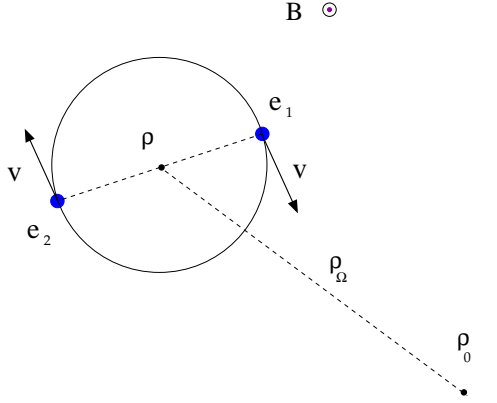}
\caption{\label{Con1} \small
Configuration I:\
Both particles rotate symmetrically on the same circle of diameter $\rho$
with frequency $\om$.
If $e_c\neq 0$, the frequency $\om=\frac{(e_1-e_2)}{m_1-m_2}B$ and the center
of the orbit is fixed in the space.
For identical particles $(e>0, m)$ for a given $\rho$ the frequency takes two values
$\om = \frac{e\,B}{2\,m}(1 \pm \sqrt{1-\frac{8\,m}{B^2\,\rho^3}})$
corresponding to different $\rm v$;
the center of the orbit rotates around the guiding center $\boldsymbol \rho_{0}$ with frequency $\Om = \frac{e\,B}{m}$, see text.
}
\end{center}
\end{figure}

The special circular trajectory is called of the first type (Configuration I) corresponds to two charges moving on the same circle, shifted to phase $\pi$, making clockwise (or counterclockwise) rotation, see Fig.~\ref{Con1}. This trajectory emerges either for two different charges,
\begin{equation}
\label{CaseI-1}
    e_1\neq e_2\ ,\ m_1\neq m_2\ ,
\end{equation}
or, for identical charges
\begin{equation}
\label{CaseI-2}
    e_1\ =\ e_2\ ,\ m_1\ =\ m_2\ .
\end{equation}
It is evident that for such a trajectory the initial velocities should be collinear, antiparallel and equal
\begin{equation}
\label{CaseI}
    \rm{v}_1=-\rm{v}_2 \equiv v\ .
\end{equation}

\bigskip

(i) {\it Different charges}

\bigskip

Let us assume two different charges (\ref{CaseI-1}). The constraints (\ref{constrain1}) - (\ref{constrain2}) can be written as
\begin{equation}
\begin{aligned}
   &{\rm v} = \frac{B\,(e_1-e_2)}{2\,(m_1-m_2)}\,\rho \ ,\\
   &\rho^3 = \frac{2\,{(m_1-m_2)}^2e_1\,e_2}{B^2\,(e_2-e_1)\,M\,e_c}\ .
\end{aligned}
\label{config1}
\end{equation}
The frequency is fixed and defined by a magnetic field $B$. For a given magnetic field $B$
there exists a single value of the initial velocity for which this trajectory occurs with $\rho$ given by (\ref{config1}). If masses are equal, $m_1 = m_2$ there is no such a trajectory. In particular, for electron-positron pair such a trajectory does not exist.

\hskip 1cm
A general situation is described by four particular cases corresponding to (counter)clockwise rotation
\footnote{The same circular trajectory occurs if $(e_1 \rar -e_1, e_2 \rar -e_2,
          m_1\lrar m_2$)\,.}
:

\bigskip

\textbf{(a)}  $q\neq 0\ ,\ e_c\neq 0$ and $\rm v>0$ (clockwise rotation)\ ,
$e_1>0,\quad e_2>0$\ ,

\begin{equation}
 \frac{m_1}{m_2} > \frac{e_1}{e_2} >1 \ .
\label{config1a}
\end{equation}
or
\begin{equation}
 \frac{m_1}{m_2} <\frac{e_1}{e_2}<1
\end{equation}

\textbf{(b)}  negative $\rm v<0$ (counterclockwise rotation),\
$e_1<0\ ,\quad e_2<0$\ ,

\begin{equation}
\quad \frac{m_1}{m_2} > \frac{e_1}{e_2} >1 \ .
\label{config1as}
\end{equation}
or
\begin{equation}
\quad \frac{m_1}{m_2} < \frac{e_1}{e_2} < 1 \ .
\end{equation}

\textbf{(c)} $e_c\neq 0$ and $\rm v>0$ (clockwise rotation)

\begin{equation}
\begin{aligned}
e_1>0,\quad e_2<0,\quad m_1 > m_2  \ .
\end{aligned}
\label{config1b}
\end{equation}

\textbf{(d)} negative $\rm v<0$ (counterclockwise rotation)

\begin{equation}
\begin{aligned}
e_1<0,\quad e_2>0,\quad m_2 > m_1  \ .
\end{aligned}
\label{config1bs}
\end{equation}

If none of conditions (\ref{config1a})-(\ref{config1bs}) is fulfilled, the circular trajectory of Fig.~\ref{Con1} does not occur. For both cases (c)-(d) the neutral system $q=0$ (the rotating dipole) appears as a particular case.
It is worth noting that the trajectory of Fig.1 can occur even in the case of particles of opposite charges.

In general, for this trajectory of Fig.1 the pseudomomentum (\ref{pseudomomentum}) (when defined with respect to the center of the trajectory) vanishes,
\begin{equation}
   K \ =\ 0 \ .
\label{K1}
\end{equation}
The total \emph{canonical} momentum $L^{total}_z\ =\ L_z + \ell_z$ (see (\ref{LzT2})) is conserved. Although, in general, none of the two terms $L_z,\ \ell_z$ \footnote{the center of the circular trajectory is chosen as the reference point\,.} is conserved separately, for the Configuration I the quantities $L_z$ and $\ell_z$ are both constants of motion (in spite of absence of separation of variables in center-of-mass-system):
the Poisson bracket $L_z$ ($\ell_z$) with Hamiltonian vanishes on a circular trajectory (\ref{traject}), see Fig.1. Hence, the quantities (${\cal { H}},\,\boldsymbol K^2,\,L^{total}_z,\,\ell_z$) are constants of motion, they span commutative Poisson algebra on the circular trajectory. Thus, on this trajectory the system is completely-integrable.
Taking into account that $K_x$ and $I\ =\ (\boldsymbol\rho \cdot \boldsymbol p)$,
where $\boldsymbol p$ is the canonical relative momentum (\ref{CMvar}), are also conserved the system becomes super-integrable.

%\pagebreak

\bigskip

(ii) {\it Rotating dipole}

\vspace{0.2cm}

Consider a rotating dipole $e_1=-e_2 \equiv e>0$ ($q=0$) with  $m_1 > m_2$, see
the Case (c). From (\ref{config1}) it follows that
\begin{equation}
\begin{aligned}
& {\rm v} = e{\bigg[ \frac{B}{M\,(m_1-m_2)}\bigg]}^{\frac{1}{3}}\,,
\qquad \rho = \bigg[ \frac{{(m_1-m_2)}^{2}}{B^{2} M }\bigg]^{\frac{1}{3}}\ .
\end{aligned}
\label{neutral1a}
\end{equation}
In this case the pseudomomentum $K$ vanishes, see (\ref{K1})
\footnote{It is worth noting that for a neutral system, $q=0$, the Pseudomomentum does not depend on the reference point and always vanishes (see (\ref{K1}))\,.}.
The total Hamiltonian and the total angular momentum take values
\[ {\cal H}\ =\ \frac{e^2\,M}{2}
{\bigg(\frac{B}{m_1^2-m_2^2}\bigg)}^{\frac{2}{3}}-\frac{e^2}{\rho}\ .
\]
\[
L_z^{total}\ = \ \frac{e\ (m_2^2-m_1^2)^{\frac{1}{3}}}
   {2 \ B^{\frac{1}{3}}\ M\ }    \ ,
\]
respectively.
On the trajectory of Fig.1 the relative angular momentum $\ell_z$ is equal to
\begin{equation}
 \ell_z = \frac{e\ (m_2-m_1)^{\frac{1}{3}}(m_1^2+6\,m_1\,m_2+m_2^2)}
   {4 \ B^{\frac{1}{3}}\,M^{\frac{5}{3}}}    \ .
%\mathbf {\hat z} \ ,
\label{LR1q}
\end{equation}

%\pagebreak

(iii) {\it Identical particles}

\vspace{0.2cm}

In the case of two identical particles, $m_1=m_2\equiv m$, $e_1=e_2\equiv e>0$\  (\ref{CaseI-2}) the coupling charge vanishes, $e_c=0$.
The constraints (\ref{constrain1})-(\ref{constrain2}) do not emerge in this case: the Newton equations (\ref{NEsep}) for the particle $1$ and $2$ do simply coincide. The circular trajectory of the diameter $\rho$ appears \cite{CC:1997} being characterized by
\begin{equation}
  {\rm v}=\frac{e\,B\,\rho}{4\,m}\bigg(1\pm \sqrt{1-\frac{8\,m}{B^2\,\rho^3}}\bigg)\,.
\label{ec1}
\end{equation}
Thus, for given $\rho$ there exist two different initial velocities leading to the same circular trajectory. Howover, for given ${\rm v}$ there exists a single circular trajectory with a certain $\rho$. It corresponds to clockwise rotation with frequency equal to $\frac{2 \,{\rm v}}{\rho}$. It is interesting that for a given magnetic field there exists a minimal circular trajectory with $\rho_{min}={(\frac{8\,m}{B^2})}^{\frac{1}{3}}$ (it corresponds to vanishing the square root in (\ref{ec1})).

\hskip 1cm
In general, for $e_c=0$ (when charges have equal cyclotron
frequencies) the separation of variables occurs
\begin{equation}
\begin{aligned}
{\cal { H}}& = {\cal {H}}_{R}(\mathbf {P},\mathbf R)+{\cal {H}}_{\rho}
   (\mathbf {p},\boldsymbol \rho)\\
& \equiv \bigg[\frac{ {( \mathbf {P}-\frac{e\,M}{m_1}\,{\mathbf A}_{\mathbf R} )}^2}{2\,M}\bigg]
+ \bigg[\frac{ {( \mathbf {p}-\frac{e\,m_2}{M} {\mathbf A}_{\boldsymbol \rho} )}^2}{2\,m_r}+\frac{m_2}{m_1}\,\frac{e^2}{\rho}\bigg]\ ,
\end{aligned}
\end{equation}
the canonical transformation (\ref{CanTrans}) becomes the identity transformation. The term ${\cal {H}}_{R}$ is the Hamiltonian of a composite particle of charge $q$ and mass $M$ in the presence of a constant magnetic field.

\vspace{0.2cm}

The circular trajectory can be easily found
\begin{equation}
{\boldsymbol \rho}_{1} =  \boldsymbol \rho_\om + \boldsymbol \rho_\Om +
\boldsymbol \rho_0\ ,
\qquad {\boldsymbol \rho}_{2} = -\boldsymbol \rho_\om + \boldsymbol \rho_\Om +
\boldsymbol \rho_0\ ,
\end{equation}
with
\begin{equation}
\begin{aligned}
&  {\boldsymbol \rho_\om} = \frac{\rm{v}}{\om}\,(\cos \om t,\, -\sin \om t)\,,
\qquad \,\,\,\, {\om} = \frac{e\,B}{2\,m}\bigg(1\pm \sqrt{1-\frac{8\,m}{B^2\,\rho^3}}\bigg)\,,
\\ & {\boldsymbol \rho_\Om} =  \frac{{V}_R}{\Om} \,(\cos \Om t,\, -\sin \Om t)\,,
\qquad {\Om} = \frac{e\,B}{m}\,,
\end{aligned}
\end{equation}
(cf.(\ref{traject})),
where ${\boldsymbol \rho_0}$ is an arbitrary constant vector and $V_R$ is a parameter. A physical meaning of this solution is the following.
The center of the circular trajectory of the relative motion (see Fig.~\ref{Con1}) moves with velocity $V_R$ making rotation with frequency $\Om=\frac{e \,B}{m}$ around to the
fixed point $\boldsymbol \rho_0$ which plays the role of the guiding center for the whole system. If $V_R$ tends to zero, then the center of the circular
trajectory (relative motion) tends to $\boldsymbol \rho_0$.

\hskip 1cm
In general, for $e_c=0$ the Pseudomomentum does not vanish, $\boldsymbol K\neq 0$. However, for a particular case of identical particles a relation (\ref{K2}) appears, which connects
$\boldsymbol K^2$ with $L_z$ and ${\cal {H}}_{R}$, $L_z$ and $\ell_z$ are conserved. It can be shown if the guiding center
\begin{equation}
\begin{aligned}
\boldsymbol { \rho}_{0}=\frac{ {\boldsymbol {K}}\times \mathbf{B} }{ 2 \,e\,B^2}\ ,
\end{aligned}
\label{rho0R}
\end{equation}
\vspace{0.2cm}
is chosen as a reference point the Pseudomomentum vanishes, $K=0$.
It can be checked that the following four quantities (${\cal {H}}_{R},\,{\cal {H}}_{\rho},\,L_z,\,\ell_z$) with the guiding center $\boldsymbol {\rho}_{0}$ chosen as the reference point are constants of motion for a circular trajectory,
\begin{equation}
\begin{aligned}
&{\cal {H}}_{R} = m \,V_R^2 \ ,
\\ & \boldsymbol L_{z} = -\frac{m^2\,V^2_R}{e\,B}\,{\mathbf {\hat z}}\ ,
\\ & {\cal {H}}_{\rho} = \frac{e^2\,B^2\,\rho^2}{16\,m}{\bigg(1\pm \sqrt{1-\frac{8\,m}{B^2\,\rho^3}}\bigg)}^2 + \frac{e^2}{\rho}\ ,
\\ & \boldsymbol \ell_{z} =  \mp \frac{e\,B\,\rho^2}{4}\sqrt{1-\frac{8\,m}{B^2\,\rho^3}}                     \,{\mathbf {\hat z}} \ .
\end{aligned}
\end{equation}
while $\boldsymbol K^2$ is given by (\ref{K2}). ${\cal {H}}_{R},\,{\cal {H}}_{\rho},\,L_z,\,\ell_z$ span commutative Poisson algebra, where ${\cal {H}}_{\rho},\ \ell_z$ are particular integrals.
For $e_c=0$ the system is completely-integrable globally, the global integrals take values (35) on a circular trajectory.

%For a circular trajectory any function $f=f({\rm v},\,\rho,\,\boldsymbol \rho \cdot {\mathbf v})$ with $\mathbf v$ written in terms of canonical variables,
%$\mathbf v = \frac{1}{m_r}(\mathbf p - \frac{e\,m_2}{M}\mathbf A_{\boldsymbol \rho})$
%is a constant of motion, $\{f,\,{\cal {H}} \}=0$.

%\clearpage

%\begin{center}
\subsection{Configuration II}
%\end{center}

\bigskip

\begin{figure}
\begin{center}
\includegraphics[width=2.5in,angle=0]{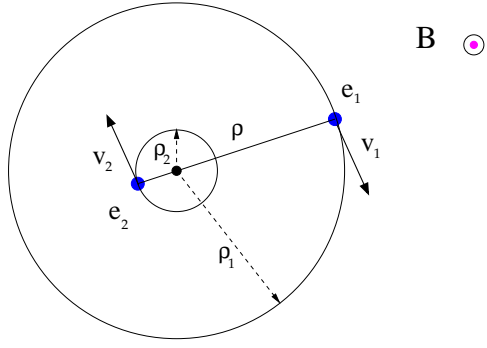}
\caption{\label{Con2} \small Both particles rotate on concentric circles with relative phase $\pi$ with angular frequency $\om$.
If $e_c=0$ the frequency $\om = \frac{e_1\,B}{2\,m_1}(1\pm \sqrt{1-\frac{4\,m_2}{B^2\,\rho_1\,\rho^2}})$\ ,\ if $e_c\neq 0$ the
frequency $\om = \frac{B\,(e_1\,{\rm v}_1+e_2\,{\rm v}_2)}{m_1\,{\rm v}_1+m_2\,{\rm v}_2}$.}
\end{center}
\end{figure}

The special circular trajectory is called of the second type (Configuration II) when two charges move on two concentric circles with relative phase $\pi$, see Fig.~\ref{Con2}.
For such a trajectory the initial velocities should be chosen collinear and antiparallel,
\begin{equation}
\label{CaseII}
    \rm{v}_1 \neq -\rm{v}_2\ ,\ v_1 v_2 < 0\ ,
\end{equation}
with the assumption
\[
    |\rm{v}_1| > |\rm{v}_2| \ ,
\]
that implies the first particle moves on the external circle.

If $\rm v_1 > 0$ the clockwise rotation occurs with $\rm v_2 < 0$, otherwise $\rm v_1 < 0$ corresponds to the counterclockwise rotation with $\rm v_2 > 0$. Analyzing the constraints (\ref{constrain1}) - (\ref{constrain2}) one can enlist the pairs of charges for which the Configuration 2 appears.

1.\ {\it Clockwise rotation}, ${\rm v}_1>0$\ .

If $$ {\rm v}_1>\frac{m_2}{m_1}\,|{\rm v}_2| \ ,$$ we have the following cases:

\bigskip

 \textbf{(1a)} charges of opposite sign

$$e_1>0\ ,\ e_2<0\,.$$

 \textbf{(1b)} charges of the same sign

 $$e_{1,2}>0\ ,\ \frac{e_1}{m_1}<\frac{e_2}{m_2}\ ,\ {\rm v}_1 >
 \frac{e_2}{e_1}\,|{\rm v}_2| \ ,$$

or

 $$e_{1,2}<0\ ,\ \frac{e_1}{m_1}<\frac{e_2}{m_2}\ ,\ {\rm v}_1 <
 \frac{e_2}{e_1}\,|{\rm v}_2| \ . $$

If $$ {\rm v}_1<\frac{m_2}{m_1} \,|{\rm v}_2| \ ,$$
we have the following cases:

\bigskip

 \textbf{(1c)} charges of opposite sign

$$e_1<0\ ,\ e_2>0\,.$$

 \textbf{(1d)} charges of the same sign

 $$e_{1,2}>0\ ,\ \frac{e_1}{m_1}>\frac{e_2}{m_2}\ ,\ {\rm v}_1 <
 \frac{e_2}{e_1}\,|{\rm v}_2| \,, $$ or

 $$e_{1,2}<0\ ,\ \frac{e_1}{m_1}>\frac{e_2}{m_2}\ ,\ {\rm v}_1 >
 \frac{e_2}{e_1}\,|{\rm v}_2| \,.      $$

\bigskip

If $$ {\rm v}_1\ =\ \frac{m_2}{m_1} \,|{\rm v}_2| \ ,$$
we have the case $e_c=0$.

\bigskip

2.\ {\it Counterclockwise rotation}, ${\rm v}_1<0$\ .

If $$ {\rm v}_2>\frac{m_1}{m_2}\,|{\rm v}_1| \ ,$$ we have the following cases:

\bigskip

 \textbf{(2e)} charges of opposite sign

$$e_1>0\ ,\ e_2<0\,.$$

 \textbf{(2f)} charges of the same sign

 $$e_{1,2}>0\ ,\ \frac{e_1}{m_1}<\frac{e_2}{m_2}\ ,\ {\rm v}_2 <
 \frac{e_1}{e_2}\,|{\rm v}_1| \,, $$

or

 $$e_{1,2}<0\ ,\ \frac{e_1}{m_1}<\frac{e_2}{m_2}\ ,\ {\rm v}_2 >
 \frac{e_1}{e_2}\,|{\rm v}_1| \,. $$

If $$  {\rm v}_2<\frac{m_1}{m_2}\,|{\rm v}_1| \ ,$$ we have the cases

\bigskip

 \textbf{(2g)} charges of opposite sign

$$e_1<0\ ,\ e_2>0\,.$$

 \textbf{(2h)} charges of the same sign

$$e_{1,2}>0\ ,\ \frac{e_1}{m_1}>\frac{e_2}{m_2}\ ,\ {\rm v}_2 >
\frac{e_1}{e_2}\,|{\rm v}_1| \ , $$ or

$$e_{1,2}<0\ ,\ \frac{e_1}{m_1}>\frac{e_2}{m_2}\ ,\ {\rm v}_2 <
\frac{e_1}{e_2}\,|{\rm v}_1| \ . $$

If $$ {\rm v}_1\ =\ \frac{m_2}{m_1} \,|{\rm v}_2| \ ,$$
we have the case $e_c=0$.

\bigskip

\hskip 1cm
Therefore, a trajectory of the second type exists for a pair of
arbitrary charges $(e_1, e_2)$. For masses $(m_1, m_2)$ and the initial velocities $\rm v_1>0\ ,\ \rm v_2<0$, which obey a condition
\[
   \frac{(m_1\,{\rm v_1}+m_2\,{\rm v_2})\,e_c}{e_1\,e_2}<0\ ,
\]
one can always indicate a magnetic field
\[
B \ = \ \frac{{(\,{\rm v_1}-{\rm v_2})}^2(m_1\,{\rm v_1}+m_2\,{\rm v_2})\,M\,e_c\,{\rm v_1}\,{\rm v_2}}{e_1\,e_2\,{(e_1\,{\rm v_1}+e_2\,{\rm v_2})}^2}  \]
for which a circular trajectory appears (see Fig.~\ref{Con2}) of the diameter
\begin{equation}
\rho\ = \ \frac{e_1\,e_2\,{(e_1\,{\rm v_1}+e_2\,{\rm v_2})}}{{(\,{\rm v_1}-{\rm v_2})}\,M\,e_c\,{\rm v_1}\,{\rm v_2}} >0 \ .
\label{con2g}
\end{equation}
It can be shown that on a trajectory of the second type the following expressions
\begin{equation}
\begin{aligned}
&{\cal {H}}_{R} = \frac{ {(m_1\,{\rm v}_1+m_2\,{\rm v}_2)}^2}{2\,M} \,, \\
& \boldsymbol L_{z} = -\frac{{(m_1\,{\rm v}_1+m_2\,{\rm v}_2)}^3}{2\,B\,M\,(e_1\
      {\rm v}_1+e_2\,{\rm v}_2)}\,{\mathbf {\hat z}}\ ,\\
& {\cal {H}}_{\rho} =\frac{m_r \,{({\rm v}_1-{\rm v}_2)}^2}{2} +\frac{e_1\,e_2}{\rho}\ ,\\
& \boldsymbol \ell_{z} = \frac{({\rm v}_2 - {\rm v}_1)(m_1\
   {\rm v}_1+m_2\,{\rm v}_2)}{2\, B\, M\ (e_1\,{\rm v}_1+e_2\,{\rm v}_2)^2}
   \bigg[ e_2 \,m_1\, {\rm v}_2\,(m_1\,{\rm v}_1 + 2\, m_2\,{\rm v}_1\, - m_2\,{\rm v}_2)  \\
& + e_1 \,m_2\, {\rm v}_1\ (m_1\,{\rm v}_1 - 2\, m_1\,{\rm v}_2\, - m_2\,{\rm v}_2)\ \bigg]{\mathbf {\hat z}} \ .
\end{aligned}
\label{integrals}
\end{equation}
are constants of motion. Functions ${\cal {H}}_{\rho}, \boldsymbol \ell_{z}$ are independent particular integrals.  On this trajectory the Pseudomomentum $K=0$  (see (\ref{K1})) as well as $I\ =\ (\boldsymbol\rho \cdot \boldsymbol p)=0$. The trajectory is super-integrable.

\hskip 1cm
We focus on two particular cases, $q=0$ (rotating dipole) and $e_c=0$ (the charges of the same sign with equal cyclotron frequencies).

\bigskip

$\bullet$\quad \textbf{Case $q=0$ (rotating dipole)}

\vspace{0.3cm}

Let $e_1=-e_2=e$. In addition to the constraints (\ref{constrain1})-(\ref{constrain2}) for arbitrary $B>0$, it is naturally assumed that
\begin{equation}
\rho =  \frac{ m_1\,\text{v}_1+m_2\,\text{v}_2 }{ e_1\,B }>0 \ ,
\label{q2-1}
\end{equation}
as well as a reality of
\begin{equation}
 \text{v}_1 = -\frac{m_2\, \text{v}_2}{2\,m_1}\bigg ( 1 \pm \sqrt{1-\frac{4\,e^3\,B\,m_1}{M\,m_2^2\,\text{v}_2^3}}\bigg)\equiv \text{v}^{(\pm)}_1
 \ .
\label{q2-2}
\end{equation}
it implies the expression under the square root must be non negative
\begin{equation}
   \frac{4\,e^3\,B\,m_1}{M\,m_2^2\,\text{v}_2^3} < 1\ .
\label{q2-3}
\end{equation}
A simple analysis leads to four particular situations
\begin{equation}
\begin{aligned}
   &\text{v}_2>0,\, e > 0     \, \Rightarrow  \, \text{v}^{(\pm)}_1<0 \ \
   \mbox{(counterclockwise rotation, two solutions)}\ ,
\\ &\text{v}_2>0,\, e < 0  \, \Rightarrow  \, \text{v}^{+}_1<0 \ \
   \mbox{(counterclockwise rotation, one solution)}\ ,
\\ &\text{v}_2<0,\, e > 0  \, \Rightarrow  \, \text{v}^{+}_1>0 \ \
   \mbox{(clockwise rotation, one solution)}\ ,
\\ &\text{v}_2<0,\, e < 0  \, \Rightarrow  \, \text{v}^{(\pm)}_1>0 \ \
   \mbox{(clockwise rotation, two solutions)}\ .
\end{aligned}
\label{v1con2}
\end{equation}
If parameters of the system and initial data satisfy the constraints (\ref{q2-1})-(\ref{q2-3}), the circular trajectories shown in Fig.~\ref{Con2} appear. Putting $\rm v_1 = -\rm v_2$ we recover (\ref{neutral1a}).

\hskip 1cm
For the Configuration II in the case $q=0$ the cms angular momentum $L_z$ and relative angular momentum $\ell_z$ are conserved separately in spite of the absence of separation of variables(!): the Poisson brackets with Hamiltonian vanish. They take values \footnote{The center of the circular trajectory is chosen as the reference point\,.}
\begin{equation}
\begin{aligned}
&{\cal {H}} = \frac {{(m_1\,{\rm v}_1+m_2\,{\rm v}_2)}^2}{2\,M} +\frac{m_r\,{({\rm v}_1-{\rm v}_2)}^2}{2}-\frac{e^2}{\rho} \ ,\\
& K=0\ ,\\
& \boldsymbol L_{z} = \frac{{(m_1\,{\rm v}_1+m_2\,{\rm v}_2)}^3}{2\,e\,B\,M
({\rm v}_2-{\rm v}_1)} \ {\mathbf {\hat z}}\ ,\\
& \boldsymbol \ell_{z} =  \frac{{(m_1\,{\rm v}_1+m_2\,{\rm v}_2)}
  [(m_1^2+m_2^2){\rm v}_1{\rm v}_2-m_1\,m_2 ({\rm v}_1^2+{\rm v}_2^2-4\,{\rm v}_1\,{\rm v}_2) ]}{2\,e\,B\,M({\rm v}_1-{\rm v}_2)}\ {\mathbf {\hat z}} \ ,
\end{aligned}
\label{Int-IIq}
\end{equation}
(cf. (\ref{integrals})) with ${\rm v}_1\,,\ \rho$ from (\ref{q2-1})-(\ref{q2-2}). Hence, the quantities $({\cal { H}}\ ,\ \boldsymbol K^2\ ,\ L_z\ ,\ \ell_z)$ are constants of motion, all of them are in involution. On the trajectories of the second type the system is completely-integrable.
${\cal {H}}_{\rho},\ I\ =\ (\boldsymbol\rho \cdot \boldsymbol p)$ are extra particular integrals. The trajectory is super-integrable. Thus, one can say that for $q=0$ the system is particularly integrable \cite{Turbiner:2012}.

\bigskip

$\bullet$\quad \textbf{Case $e_c=0$}

\vspace{0.3cm}

In this case the Newton equations (\ref{NEsep}) coincide.
It leads to the constraints
\begin{equation}
\begin{aligned}
& \frac{\text{v}_1}{\text{v}_2}\ =\ -\frac{m_2}{m_1}   \ ,
\\ & \text{v}_1 \ =\ \frac{e_1\,B\,\rho_1}{2\,m_1}\bigg(1\pm \sqrt{1-\frac{4\,m_2}{B^2\,\rho_1\,\rho^2}}\bigg)   \ ,
\\ &\om \ =\ \frac{\text{v}_1}{\rho_1}=-\frac{\text{v}_2}{\rho_2} \ .
\end{aligned}
\label{ec2}
\end{equation}
If initial data obey these constraints, the circular trajectories of the second type appear, see Fig.~\ref{Con2}. It is worth noting that for given $\rho_1,\ \rho_2$ there exist two different initial velocities $\text{v}_1$ leading to the same circular trajectory.
%For a given magnetic field $B$ there exists a minimal  $\rho_{1,min}=\bigg[{\frac{4\,m_2\,\text{v}_1^2}{B^2\ {(\text{v}_1-\text{v}_2)}^2}}\bigg]^{\frac{1}{3}}$.
In this case the corresponding individual trajectories are
\begin{equation}
{\boldsymbol \rho}_{1} =  \boldsymbol \rho_{\om,1} + \boldsymbol \rho_\Om +\boldsymbol \rho_0\,,
\qquad {\boldsymbol \rho}_{2} = \boldsymbol \rho_{\om,2} + \boldsymbol \rho_\Om +\boldsymbol \rho_0\,,
\end{equation}
with
\begin{equation}
\begin{aligned}
&  {\boldsymbol \rho_{\om,1(2)}} = \frac{\rm{v}_{1(2)}}{\om}\,(\cos \om t,\, -\sin \om t)\,,
\\ & {\boldsymbol \rho_\Om} =  \frac{{V}_R}{\Om} \,(\cos \Om t,\, -\sin \Om t)\,,
\\ & {\Om} = \frac{e_1\,B}{m_1}\,,
\end{aligned}
\end{equation}
where $\boldsymbol \rho_0$ is a constant vector, ${V}_R$ is a parameter, $\om$ is defined in (\ref{ec2}).
If $m_1=m_2=m$ then $\text{v}_1 = -\text{v}_2 = \text{v} > 0$ we arrive at the trajectory of the first type (Configuration I) for identical particles with ${\rho_1}=\frac{\rho}{2}$ and recover (\ref{ec1}).

\hskip 1cm
It can be easily checked that four quantities
\footnote{The guiding center
$\boldsymbol {\rho}_{0}$ is chosen as the reference point\,.}
\begin{equation}
\begin{aligned}
&{\cal {H}}_{R} = \frac{ {M\,V_R^2}}{2} \ ,\\
& \boldsymbol L_{z} = -\frac{m_1\,M\,V_R^2}{2 e_1 B}\
  {\mathbf {\hat z}}\ ,\\
& {\cal {H}}_{\rho} =\frac{m_r \,{({\rm v}_1-{\rm v}_2)}^2}{2} +\frac{m_2}{m_1}
  \frac{e_1^2}{\rho}\ ,\\
& \boldsymbol \ell_{z} =  \mp \frac{M\,e_1\,B\,{\rho_1}^2\,}
   {2 \,m_2}\sqrt{1-\frac{4\,m_2}{B^2\,\rho_1\,\rho^2}}\ {\mathbf {\hat z}} \ ,
\end{aligned}
\label{integral-IIec}
\end{equation}
are constants of motion in involution forming a commutative Poisson algebra.
It holds on a level of the Poisson brackets. Hence, for $e_c=0$ the system is completely-integrable.
$I\ =\ (\boldsymbol\rho \cdot \boldsymbol p)=0$ is extra particular integral.

%It appears one more particular integral which looks like as the relative Pseudomomentum
%$ {\boldsymbol k}_{\rho}\ =\ {\mathbf p} + e_{\rho}\,{\mathbf A}_{\boldsymbol \rho}$
%with an effective charge $e_{\rho}$ is a constant on the circular trajectory of the second type at $e_c=0$.

\subsection{Configuration III}

\bigskip

\begin{figure}
\begin{center}
\includegraphics[width=2.5in,angle=0]{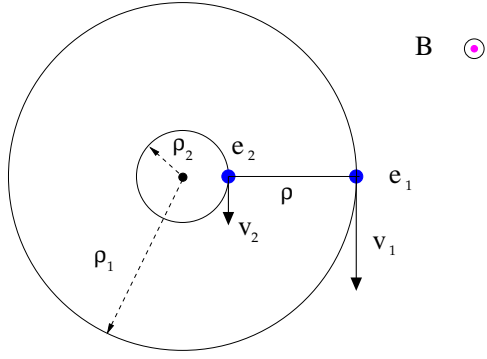}
\caption{\label{Con3} \small Both particles rotate with the zero relative phase on concentric circles with angular frequency
$\om \ =\ \frac{B\,(e_1\,{\rm v}_1+e_2\,{\rm v}_2)}{m_1\,{\rm v}_1+m_2\,{\rm v}_2}$.}
\end{center}
\end{figure}

The special circular trajectories are called of the third type (Configuration III) when two charges move on two concentric circles with zero relative phase, see Fig.~\ref{Con3} (cf. Fig.~\ref{Con2}).
It is evident that for such a trajectory the initial velocities should be chosen collinear and parallel, they obey the constraints
\begin{equation}
 \quad {\rm v}_1 > {\rm v}_2 > 0 \quad , \quad \frac{m_2}{e_2} > \frac{m_1}{e_1}\ ,
\label{CaseIIIco}
\end{equation}
for counterclockwise rotation and
\begin{equation}
 \quad {\rm v}_1 < {\rm v}_2 < 0 \quad , \quad \frac{m_2}{e_2} < \frac{m_1}{e_1}\ ,
\label{CaseIIIcl}
\end{equation}
for clockwise rotation. For both rotations, the condition
\begin{equation}
   e_1\,{\rm v}_1 > -e_2\,{\rm v}_2 \ ,
\label{CaseIIIconst}
\end{equation}
has to be fulfilled. By definition above implies that the first charge always
moves on the external circle.
Such a circular trajectory (see Fig.~\ref{Con3}) appears in a magnetic field
\[
   B \ = \ \frac{{(\,{\rm v_1}-{\rm v_2})}^2 (m_1\,
    {\rm v_1}+m_2\,{\rm v_2})\,M\,e_c\,{\rm v_1}\,{\rm v_2}}
   {e_1\,e_2\,{(e_1\,{\rm v_1}+e_2\,{\rm v_2})}^2}\ >\ 0  \ ,
\]
and is characterized by a relative distance between charges
\[
\rho\ = \ \frac{e_1\,e_2\,{(e_1\,{\rm v_1}+e_2\,{\rm v_2})}}{{(\,{\rm v_1}-{\rm v_2})}\,M\,e_c\,{\rm v_1}\,{\rm v_2}} \ >\ 0 \ .
\]

In principle, the conditions (\ref{CaseIIIco}) - (\ref{CaseIIIconst}) are not very restrictive, for almost any charges $e_1$ and $e_2$ one can indicate their masses $m_1$ and $m_2$ for which they are satisfied. However, in two important particular cases of equal masses there is no solutions when $q=0$ and $e_c=0$.

Four constants of motion for this trajectory are given by (\ref{integrals}). The pseudomomentum vanishes, $K=0$ on this trajectory. $I\ =\ (\boldsymbol\rho \cdot \boldsymbol p)=0$ is a particular integral.

$\bullet$\quad \textbf{Case $e_1=-1, e_2=2$ }

If $e_1=-1\, , \, e_2=2$, the circular trajectories of the third type (see Fig.~\ref{Con3}) exist for
\[
 \rm v_1 > \rm v_2> \frac{\rm v_1}{2} > 0 \ ,
\]
(cf.(\ref{CaseIIIco})) for arbitrary masses. For chosen initial velocities these trajectories appear for the magnetic field
\[
  B \ = \ \frac{{(\,{\rm v_1}-{\rm v_2})}^2\,(m_1\,{\rm v_1}+m_2\,{\rm v_2})(2\,m_1+m_2)\,{\rm v_1}\,{\rm v_2}}{2{(2\,{\rm v_2}-\,{\rm v_1})}^2} \ ,
\]
with the relative distance
\[
  \rho\ = \ \frac{2\,{(2\,{\rm v_1}-\,{\rm v_2})}}{{(\,{\rm v_1}-{\rm v_2})}\,(2\,m_1+m_2)\,{\rm v_1}\,{\rm v_2}} \ .
\]

Concluding we have to emphasize that the Configuration III is more restrictive than the Configuration II which appear for arbitrary charges $e_1, e_2$. In particular,
the Configuration III does not appear for both neutral system $q=0$ and $e_c=0$.

\subsection{Configuration IV}

\begin{figure}[h]
\begin{center}
\includegraphics[width=1.5in,angle=0]{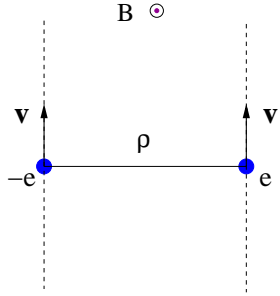}
\caption{ \small A dipole with fixed dipole moment moves with constant velocity.}
\label{Case4}
\end{center}
\end{figure}

Let us consider a neutral system ($e_1=-e_2=e>0$) with collinear and equal initial velocities $\mathbf v_1=\mathbf v_2 \equiv\,\mathbf{v}$ and impose constraints
\begin{equation}
 (\mathbf v, \boldsymbol \rho)=0\ \mbox{and} \quad\rho=\sqrt{\frac{e}{B\, \text{v}}}\ .
\label{constrain3}
\end{equation}
In this case the Lorentz force acting on charge 1(2) is equal to the Coulomb force (see eqs.(\ref{NEsep}), (\ref{Eq})). Hence, the constraints (\ref{constrain3}) give rise to the \emph{free} motion of a neutral system with fixed dipole moment, see Fig.~\ref{Case4}. We call this trajectory of the fourth type (Configuration IV). This free parallel motion of two charges in a constant magnetic field was first found in Curilef-Claro \cite{CC:1997}.
In this case the Pseudomomentum
\begin{equation}
\boldsymbol K=\bigg(M\,\text{v}+\sqrt{\frac{e^3\,B}{\text{v}}}\bigg)\,\mathbf{\hat v}\,,
\end{equation}
does not vanish.

For the Configuration IV the quantities $L_z$ and $\ell_z$ are conserved separately. They take the values
\footnote{The reference point is chosen such that $(\mathbf R, \boldsymbol \rho)=0$\,.}
\begin{equation}
\begin{aligned}
&{\cal {H}} = \frac {M \,{\rm v}^2}{2} - \sqrt{e^3\,{\rm v}\,B} \ ,\\
& \boldsymbol K = \bigg(M\,\text{v} + \sqrt{\frac{e^3\,B}{\text{v}}}\bigg)\
    \mathbf{\hat v}\ ,\\
& L_{z} = 0\ ,\\
& \boldsymbol \ell_{z} =  \frac{e^2\,(m_2-m_1)}{2\,M\,{\rm v}} \ {\mathbf {\hat z}} \ ,
\end{aligned}
\label{Int-IV}
\end{equation}

Hence, the quantities (${\cal { H}},\,\boldsymbol K^2,\,L^{total}_z,\,\ell_z$) are constants of motion, all of them are in involution. Here $\ell_z$ is particular integral
as well as $I\ =\ (\boldsymbol\rho \cdot \boldsymbol p)=0$. For this trajectory the system is completely-integrable being also super-integrable realizing a particular integrability \cite{Turbiner:2012}.

\subsection{Configuration V}

\begin{figure}[h]
\begin{center}
\includegraphics[width=1.5in,angle=0]{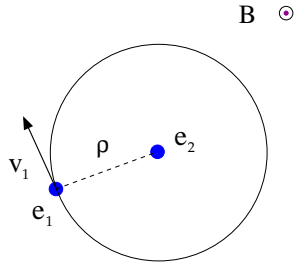}
\caption{\label{Case5} \small Circular motion in the Born-Oppenheimer approximation ($m_2 \rightarrow \infty$). }
\end{center}
\end{figure}

A circular trajectory of the fifth type (Configuration V, see Fig.~\ref{Case5}) appears when in (\ref{constrain1})-(\ref{constrain2}) both limits $\text{v}_{2}\rightarrow0$ and $m_{2} \rightarrow \infty$ are taken simultaneously. Hence, the particle 2 is standing corresponding to the fixed charged center. We call it the Born-Oppenheimer case. For a given magnetic field $B$ the trajectory of the radius $\rho > \rho_{min}$ emerges at two different values of initial velocity
\begin{equation}
\begin{aligned}
& \text{v}_1 = \frac{e_1\,B\,\rho_1}{2\,m_1}\bigg(1\pm \sqrt{1-\frac{4\,m_1\,e_2}{e_1\,B^2\,\rho_1^3}}\bigg) \,,
\\ &\om = \frac{\text{v}_1}{\rho_1}\,,\qquad \rho_1=\rho \ .
\end{aligned}
\label{BornOp}
\end{equation}
where $\rho_{min}={(\frac{4\,m_1\,e_2}{e_1\,B^2})}^{\frac{1}{3}}$ if $e_1\,e_2>0$ and
$\rho_{min}=0$ if $e_1\,e_2 < 0$. For any ${\rm v}_1$ there exists a circular trajectory
of some radius $\rho$.

The two constants of motion take the values
\begin{equation}
\begin{aligned}
& {\cal {H}}=\frac{e_1^2\,B^2\,\rho_1^2}{4\,m_1}{\bigg(1\pm
  \sqrt{1-\frac{4\,m_1\,e_2}{e_1\,B^2\,\rho_1^3}}\bigg)}^2 +\frac{e_1\,e_2}{\rho_1}\ ,\\
& \boldsymbol \ell_{1,\,z}\equiv {\boldsymbol \rho}_1\times {\mathbf p}_1=\pm
  \frac{e_1\,B\,\rho_1^2 }{2} \sqrt{1-\frac{4\,m_1\,e_2}{e_1\,B^2\,\rho_1^3}}
  \quad \mathbf {\hat z}\,.
\end{aligned}
\label{Con4}
\end{equation}

\section{Conclusions}

In the present work we give a classification of pairs of charged particles placed in a constant uniform magnetic field which admit special planar trajectories transversal to the
magnetic field direction. These trajectories represent concentric circles of finite or infinite radii. Their major characteristics is a constant relative distance $\rho$ between charges in the process of evolution. It corresponds to the appearance of a {\it particular} constant of motion
\begin{equation}
\label{integral}
    I\ =\ (\boldsymbol\rho \cdot \boldsymbol p)\ ,
\end{equation}
where $\boldsymbol p$ is the canonical relative momentum (\ref{CMvar}). $I$ is conserved and vanishes on these special trajectories (the Configurations I-V)
\begin{equation}
\label{integral-0}
    I\ =\ 0\ ,
\end{equation}
only, unlike the total Pseudomomentum $\boldsymbol K$ (\ref{pseudomomentumIND}) and the total canonical momentum ${\boldsymbol L}^{\text{total}}_z$ (\ref{LzT}) which are
conserved for {\it any} trajectory. It is worth emphasizing that for all circular trajectories
\begin{equation}
\label{integral-0-II}
    \boldsymbol K\ =\ 0\ ,
\end{equation}
(see (\ref{integrals})) and for Configuration IV
\begin{equation}
\label{integral-0-IV}
    {\boldsymbol L}_z\ =\ 0\ ,
\end{equation}
(see (\ref{Int-IV})).

There are three important particular cases for which special trajectories are admitted:

\begin{itemize}
  \item $q=0$,\ for all Configurations special trajectories occur except Configuration III.

  \item $e_c=0$,\  the special trajectories occur for Configurations I, II. For identical particles special trajectory occurs only for Configurations I.

  \item $e_1=-1\,,\,e_2=2$,\ they occurs for all Configurations except Configuration IV.
\end{itemize}

\hskip 1cm
Assuming the straightforward quantization (\ref{Hind}), (\ref{pseudomomentumIND}), (\ref{LzT}) and (\ref{integral}) by replacing the momentum
to the momentum operator one can ask whether exist eigenstates which
are common for the Hamiltonian and one of the these integrals or all of them.
Such common eigenfunctions do exist for certain discrete values of a magnetic field $B$, they can be found analytically and they are zero modes of one of these integrals \cite{ET-q}. This construction implies a certain {\it particular} integrability:
the commutator of the Hamiltonian and an operator vanishes on a subspace of the Hilbert space \cite{Turbiner:2012}.

\begin{acknowledgments}
  The authors are grateful to J. C. L\'opez Vieyra for their interest in the present work, helpful discussions and important assistance with computer calculations. A.V.T. thanks
  P.~Winternitz for a valuable remark.
  This work was supported in part by the University Program FENOMEC and by the PAPIIT
  grant {\bf IN109512} and CONACyT grant {\bf 166189}~(Mexico).
\end{acknowledgments}

\end{document}